\documentclass[article,onecolumn,superscriptaddress,nofootinbib,floatfix,onecolumn]{revtex4}

\usepackage{dcolumn}
\usepackage{bm}
\usepackage{xspace}

\usepackage[dvips]{graphicx}
\usepackage{float}
\usepackage{epsfig}
\usepackage{graphicx}
\usepackage{longtable}
\usepackage{rotating}
\usepackage{ulem}
\usepackage{subfigure}
\usepackage{amsmath,amssymb,color,hyperref}
\usepackage{txfonts}
\usepackage{mathtools}
\usepackage{tabularx}
\usepackage{booktabs}

\usepackage{soul} 

\newcommand{\mrm}[1]{\mathrm{#1}}

\def\GeV{{\text{ }\mathrm{GeV}}}
\def\MeV{{\text{ }\mathrm{MeV}}}
\def\keV{{\text{ }\mathrm{keV}}}

\def\ttt#1{\texttt{\small #1}}

\providecommand{\bbbar}{b\overline{b}}

\newcommand{\pp}{p-p}
\newcommand{\ppbar}{p-$\bar{\rm p}$}
\newcommand{\epem}{e^+e^-}
\newcommand{\alphas}{\alpha_{\rm s}}

\newcommand{\alphasmZ}{\alphas(\rm m^2_{_{\rm Z}})}
\newcommand{\alphasmW}{\alphas(\rm m^2_{_{\rm W}})}
\newcommand{\lqcd}{\Lambda_{_{\rm QCD}}}

\providecommand{\GWh}{\Gamma^{\rm W}_{\rm had}}
\providecommand{\GWhexp}{\Gamma^{\rm W}_{\rm had,exp}}

\providecommand{\GWl}{\Gamma^{\rm W}_{\rm lep}}
\providecommand{\GWhl}{\Gamma^{\rm W}_{\rm had,lep}}
\providecommand{\GWlexp}{\Gamma^{\rm W}_{\rm lep,exp}}
\providecommand{\GWtot}{\Gamma^{\rm W}_{\rm tot}}
\providecommand{\GWtotexp}{\Gamma^{\rm W}_{\rm tot,exp}}
\providecommand{\GWtotewk}{\Gamma^{\rm W}_{\rm tot,fit}}
\providecommand{\BRWh}{\rm {\cal B}^{\rm W}_{\rm had}}
\providecommand{\BRWl}{\rm {\cal B}^{\rm W}_{\rm lep}}
\providecommand{\BRWhl}{\rm {\cal B}^{\rm W}_{\rm had,lep}}
\providecommand{\BRWhexp}{\rm {\cal B}^{\rm W}_{\rm had,exp}}

\providecommand{\BRWlexp}{\rm {\cal B}^{\rm W}_{\rm lep,exp}}
\providecommand{\RW}{\rm R_\mrm{W}}
\providecommand{\RWexp}{\rm R_\mrm{W,exp}}

\newcommand{\sqrts}{\sqrt{\rm s}}
\newcommand{\Vcs}{|V_{\rm cs}|}
\newcommand{\Vcsexp}{|V_{\rm cs,exp}|}
\newcommand{\MW}{m_{_{\rm W}}}
\newcommand{\MZ}{m_{_{\rm Z}}}

\def\cO#1{{{\cal{O}}}\left(#1\right)} 

\newcommand*{\eg}{e.g.\@\xspace}
\newcommand*{\ie}{i.e.\@\xspace}

\begin{document}

\title{$\alphas$ and $\Vcs$ determination, and CKM unitarity test, from W decays at NNLO}
 
\author{David~d'Enterria}
\affiliation{CERN, EP Department, 1211 Geneva, Switzerland}
\author{Matej Srebre$^{1,}$}
\affiliation{Faculty of Mathematics and Physics, University of Ljubljana, Jadranska 19, 1000 Ljubljana, Slovenia}

\begin{abstract}
\noindent
The hadronic ($\GWh$) and total ($\GWtot$) widths of the W boson, computed at least at 
next-to-next-to-leading-order (NNLO) accuracy, are combined to derive a new precise prediction for the
hadronic W branching ratio $\BRWh~\equiv~\GWh/\GWtot$~=~$0.682 \pm 0.011_\mrm{ par}$, using the experimental
Cabibbo-Kobayashi-Maskawa (CKM) matrix elements, or $\BRWh = 0.6742 \pm 0.0002_\mrm{th} \pm 0.0001_\mrm{par}$
assuming CKM unitarity, with uncertainties dominated by the input parameters of the calculations.
Comparing the theoretical predictions and experimental measurements for various W decay observables,
the NNLO strong coupling constant at the Z pole,
$\alphasmZ = 0.117 \pm 0.042_\mrm{exp} \pm 0.004_\mrm{th} \pm 0.001_\mrm{par}$, as well as
the charm-strange CKM element, $\Vcs$~=~0.973~$\pm$~0.004$_\mrm{ exp}$~$\pm$~0.002$_\mrm{ par}$, can be extracted under
different assumptions. 
We also show that W decays provide today the most precise test of CKM unitarity for the 5 quarks
lighter than $\MW$, $\sum_\mrm{ u,c,d,s,b} |V_\mrm{ ij}|^2 = 1.999 \pm 0.008_\mrm{exp} \pm 0.001_\mrm{th}$.
Perspectives for $\alphas$ and $\Vcs$ extractions from W decays measurements at the LHC
and future $\epem$ colliders are presented.  
\end{abstract}


\maketitle

\section{Introduction}

The strong coupling $\alphas$ is one of the fundamental parameters of the Standard Model (SM), setting the
scale of the strength of the strong interaction theoretically described by Quantum Chromodynamics (QCD).
At the reference Z pole mass scale, its value amounts to $\alphasmZ$~=~0.1181~$\pm$~0.0013~\cite{PDG}
as determined from different experimental observables confronted to perturbative QCD (pQCD) calculations at
(at least) next-to-next-to-leading-order (NNLO) accuracy~\cite{alphas2015}. Given its current $\delta\alphas/\alphas
\approx 1\%$ uncertainty---orders of magnitude larger than that of the gravitational ($\rm \delta G/G\approx
10^{-5}$), Fermi ($\delta \rm G_\mrm{ F}/G_\mrm{ F}\approx 10^{-7}$), and QED ($\delta\alpha/\alpha\approx
10^{-10}$) couplings---the strong coupling is the least precisely known of all interaction strengths in
nature. 
Improving our knowledge of $\alphas$ is a prerequisite to reduce the theoretical uncertainties in the
calculations of all high-precision pQCD processes whose cross sections or decay rates depend on higher-order
powers of $\alphas$, as is the case for virtually all those measured at the LHC. In the Higgs sector, in
particular, the $\alphas$ uncertainty is currently the second major contributor (after the bottom mass) to
the parametric uncertainties of the calculations of its prevalent H$\to\bbbar$ decay, the leading one for
the H$\to c\bar{c},gg$ modes~\cite{Heinemeyer:2013tqa}, and it also introduces a 3.7\% uncertainty on
theoretical NNLO cross sections for the (dominant) Higgs production channel via gluon-gluon fusion~\cite{alphas2015}.\\

The hadronic decay widths of the electroweak bosons, $\Gamma^{\rm W,Z}_\mrm{ had}$, are high-precision
theoretical and experimental observables from which an accurate determination of $\alphas$ can be
obtained. On the one hand, the hadronic Z width --measured  with 0.1\% experimental uncertainty in
$\epem$ collisions, 
and theoretically known up to next-to-NNLO (N$^3$LO), \ie $\cO{\alphas^4}$
QCD corrections-- provides, combined with other Z-pole observables, a powerful constraint on the current $\alphas$
world average~\cite{Baak:2014ora}. On the other hand, the hadronic W width has not been used so far
in any $\alphas$ extraction. The reasons for that are twofold. First, the $\GWh$ experimental
uncertainties --of order 2\%, or 0.4\% in the case of the more precisely known $\BRWh~\equiv~\GWh/\GWtot$ branching fraction-- are
much larger than the corresponding ones for $\Gamma^{\rm Z}_\mrm{ had}$,  
whereas the $\alphas$ sensitivity of the W and Z hadronic decays comes only through small higher-order
loop corrections.  Secondly, a complete expression of $\GWh$ including all computed higher-order terms was
lacking until recently. This situation changed with the work of~\cite{KARA} that obtained
$\GWh$ including so-far missing mixed QCD+electroweak $\cO{\alphas\alpha}$ corrections, improving upon the
previous calculations of one-loop $\cO{\alphas}$ QCD and $\cO{\alpha}$ electroweak
terms~\cite{CHENG,Denner:1990tx,KNIEHL}, and two-loop $\cO{\alphas^2}$, three-loop
$\cO{\alphas^3}$~\cite{Gorishny:1990vf, Surguladze:1990tg}, and four-loop  $\cO{\alphas^4}$~\cite{BCK} QCD corrections. Despite
the progress, the work of~\cite{KARA} still contains a range of approximations (such as \eg one-loop $\alphas$
running between $\MW$ and $\MZ$, and massless quarks), 
plus no real estimation of the associated uncertainties, which hinder its use to extract $\alphas$ from a
comparison to the data.\\ 

The purpose of this letter is twofold. First, by improving upon the N$^3$LO theoretical derivation of the W hadronic
width, removing various of the approximations applied in previous works, and by combining it with the
total W decay width known at NNLO accuracy~\cite{ZFITTER,Cho:2011rk}, we obtain a theoretical 
expression of the hadronic W branching ratio with a sound determination of all associated 
uncertainties. We then compare the theoretical predictions with the experimental data, 
and thereby determine $\alphas$. 
Secondly, since the hadronic decay width is directly proportional to the sum over the first two rows of the
CKM matrix, $\sum_\mrm{ u,c,d,s,b}|V_\mrm{ ij}|^2$ (the top quark is kinematically forbidden in W decays),
we can also extract --by fixing now $\alphas$ to its current world average-- 
a precise independent value of the charm-strange quark mixing CKM element $\Vcs$, which currently has an
experimental uncertainty of $1.6\%$ ($\Vcsexp$~=~0.986~$\pm$~0.016)~\cite{PDG}. We demonstrate, at the same time,
that the measurements of W decays provide today the most stringent test of CKM matrix unitarity for all
quarks lighter than the top quark.
The developments presented here should motivate high-quality measurements of W decays using the large
datasets available at the LHC, as well as improve the $\alphas$ extraction benchmarks expected from W
measurements 
at future $\epem$ colliders such as ILC~\cite{Djouadi:2007ik}, FCC-ee~\cite{tlep}, and CEPC~\cite{CEPC}. 

\section{Hadronic W decay width at N$^3$LO accuracy}

The hadronic decay width of the W boson can be decomposed into the following contributions: 
\begin{equation}
\GWh = \vphantom{\sum_{i=1}^4}\Gamma^{(0)} +
\sum_{i=1}^4 \Gamma^{(i)}_\mrm{QCD}(\alphas^i) +
\vphantom{\sum_{i=1}^4}\Gamma_\mrm{ewk} (\alpha) +
\vphantom{\sum_{i=1}^4}\Gamma_\mrm{mixed} (\alpha \alphas)\, .
\label{eq:w_width}
\end{equation}
where $\Gamma^{(0)}$ denotes the Born decay width, $\mathcal{O}(\alphas^i)$
the higher-order QCD corrections, $\Gamma_{\mrm{ewk}}$ the electroweak corrections of order
$\mathcal{O}(\alpha)$, and $\Gamma_{\mrm{mixed}}$ the mixed electroweak+QCD corrections of
order $\mathcal{O}(\alpha \alphas)$. In the massless quark limit, the zeroth-order decay width reads
\begin{equation}
\Gamma^{(0)} = \frac{\sqrt{2} \rm G_F N_c}{12 \pi} \MW^3 \sum_\mrm{\text{quarks }i, j} |V_\mrm{ ij}|^2,\\
\label{eq:Born_mq0}
\end{equation}
where $\rm N_c$~=~3 is the number of colours, $\rm G_\mrm{F}$ is the Fermi constant, $\MW$ is the W boson
mass, and 
$|V_\mrm{ ij}|$ the CKM matrix element ij summed over quark pairs ($\rm ij=ud,us,ub,cd,cs,cb$).
The first QCD correction to the tree-level width is
\begin{equation}
\Gamma^{(1)}_{\mrm{QCD}} (\alphas) = \Gamma^{(0)} \cdot \frac{\alphas}{\pi} \, .
\label{eq:Gamma_NLO}
\end{equation}
The calculation of $\GWh$ can be factorized  
as a product of the Born width, Eq.~(\ref{eq:Born_mq0}), times the remaining terms:
\begin{equation}
\GWh = \Gamma^{(0)}\left[ 1 + \sum_{i=1}^4 c^{(i)}_\mrm{QCD}\cdot\left(\frac{\alphas}{\pi}\right)^i + \delta_\mrm{ewk}(\alpha) + \delta_\mrm{mixed}(\alpha \alphas) \right]\,,
\label{eq:GammaWh_prod}
\end{equation}
where the $c^{(i)}_\mrm{QCD}$ coefficients can be obtained from the perturbative expansion in
$\alphas$ of the well-known $e^+ e^-$ cross-section ratio $R = \frac{\sigma(e^+ e^-\rightarrow \text{
    hadrons})}{\sigma(e^+ e^- \rightarrow \mu^+ \mu^-)}$, 
calculated up to $\mathcal{O}(\alphas^4)$ in~\cite{BCK,CKK}, with coefficients (for $N_f=5$ flavours):
\begin{equation}
R = 1 + \frac{\alphas}{\pi} + 1.4097 \left(\frac{\alphas}{\pi}\right)^2  
+\left(-12.76709\right)\left(\frac{\alphas}{\pi}\right)^3 + \left(-80.0075\right)\left(\frac{\alphas}{\pi}\right)^4\,.
\label{eq:R}
\end{equation}
Numerically, the relative weights of the different partial widths 
in Eq.~(\ref{eq:w_width}) are: $\Gamma^{(0)}/\GWh\approx$~96.6\%, $\Gamma^{(1)}_\mrm{ QCD}/\GWh\approx$~3.7\%, 
$\Gamma^{(2)}_\mrm{ QCD}/\GWh\approx$~0.2\%, $\Gamma^{(3)}_\mrm{ QCD}/\GWh~\approx$~$-$0.1\%, 
$\Gamma^{(4)}_\mrm{ QCD}/\GWh\approx$~$-$0.02\%, $\Gamma_\mrm{ewk}/\GWh\approx$~$-$0.35\%, and 
$\Gamma_\mrm{\rm mixed}/\GWh\approx$~$-$0.05\%, at N$^3$LO (Table~\ref{tab:1}).
In Ref.~\cite{KARA}, the first-order QCD corrections of Eq.~(\ref{eq:GammaWh_prod}) were obtained assuming
zero quark masses, \ie directly from the coefficients of Eq.~(\ref{eq:R}), and the higher-order corrections and
renormalization constants in the QCD, electroweak and mixed terms were obtained setting the CKM matrix to unity.
Since $\Gamma^{(0)}+\Gamma^{(1)}_\mrm{ QCD}$ numerically amount to $\sim$100\% of 
$\GWh$, a first improvement over~\cite{KARA} consists in computing the exact results for the Born
width and the first QCD correction using finite quark masses, rather than through the first two coefficients
of $R$. 
In our calculations, we thus replace Eq.~(\ref{eq:Born_mq0}) with the exact expression for the 
decay width with full quark masses $m_\mrm{ q,i}$~\cite{DENNER}, namely
\begin{equation}
\Gamma^{(0)} =  \frac{\sqrt{2} \rm G_F N_c}{24 \pi} \sum_\mrm{\text{quarks }i, j} \frac{\kappa\left(\MW^2, m_\mrm{ q,i}^2, m_\mrm{ q',j}^2\right)}{\MW}
\left( 2 \MW - m_\mrm{ q,i}^2 - m_\mrm{ q',j}^2 - \frac{(m_\mrm{ q,i}^2 - m_\mrm{ q',j}^2)^2}{\MW^2} \right)
|V_\mrm{ ij}|^2\,,
\label{eq:Gamma_Whad_mq}
\end{equation}
where $\kappa(x, y, z)$ is the K\"all\'en function. Such an improved evaluation of the Born width also directly
impacts the most important QCD correction obtained through Eq.~(\ref{eq:Gamma_NLO}).
We have cross checked that our implementation of Eq.~(\ref{eq:Gamma_Whad_mq}) matches numerically the result
of Eq.~(\ref{eq:Born_mq0}) in the limit $m_\mrm{ q,i}, m_\mrm{ q',j} \rightarrow 0$, as well as the exact
leading order calculation of~\cite{CHENG}. For the remaining higher-order QCD corrections, starting from
$\mathcal{O}(\alphas^2)$, we use the coefficients given by Eq.~(\ref{eq:R}), while the electroweak and mixed
corrections are those computed in~\cite{KARA}. 
Since the main motivation of the analysis is to obtain a precise value of $\alphas$, a second direct improvement
with respect to the LO $\alphas$ expression used in~\cite{KARA} is achieved by evaluating $\alphas$ at the
relevant scales here ($\MW$ and $\MZ$) including up to three loops (\ie NNLO) in the renormalization
group $\beta$ function~\cite{Tarasov:1980au}.
Also, for our numerical evaluations we use the latest values of the SM parameters with their
associated uncertainties~\cite{PDG}: 
\begin{alignat*}{7}
  m_u &= 2.3^{+ 0.7}_{-0.5}\MeV \, ,  &\quad m_d &= 4.8^{+ 0.5}_{-0.3}\MeV \, , \notag \\
  m_c &= 1.67 \pm 0.07 \GeV \, ,  &\quad m_s &= 95 \pm 5 \MeV \, , \notag \\
  m_t &= 174.6 \pm 1.9\GeV \, ,  &\quad m_b &= 4.78 \pm 0.06 \GeV \, , \notag \\
  m_\mu &= 105.6583715 \pm 0.0000035 \MeV \, ,  &\quad m_\tau &= 1.77686 \pm 0.00012 \GeV \, , \stepcounter{equation}\tag{\theequation}\label{eq:num_input} \\
  m_H &= 125.09 \pm 0.24 \GeV \, ,  &\quad m_e &= 510.998928 \pm 0.000011 \keV \, , \notag \\
  \MW &= 80.385 \pm 0.015 \GeV \, , &\quad \MZ &= 91.1876 \pm 0.0021 \GeV \, ,	 \notag \\
  \alpha &= \left(7.2973525664 \pm 0.0000000017\right)\cdot 10^{-3}  \, ,  &\quad \rm G_F &= \left(1.1663787 \pm 0.0000006\right) \cdot 10^{-5}\GeV^{-2} \, . \notag
\end{alignat*}
Here, $m_u$, $m_d$ and $m_s$ correspond to current-quark masses, and $m_c$, $m_b$ and $m_t$ to pole masses~\cite{PDG}. 
The Higgs boson mass corresponds to the most recent LHC average value~\cite{Aad:2015zhl}.
When not left free, the QCD coupling is taken at its current world average, $\alphasmZ$~=~0.1181~$\pm$~0.0013~\cite{PDG}.
The experimental values of the CKM matrix elements used are
\begin{alignat*}{7}
|V_\mrm{ ud}| &= 0.97425 \pm 0.00022 \, ,  &\quad |V_\mrm{ cd}| &= 0.225 \pm 0.008 \, , \notag \\
|V_\mrm{ us}| &= 0.2253 \pm 0.0008 \, ,  &\quad \Vcs &= 0.986 \pm 0.016 \, ,   \stepcounter{equation}\tag{\theequation}\label{eq:PDG_CKM} \\
|V_\mrm{ ub}| &= (4.13 \pm 0.49) \cdot 10^{-3} \, ,  &\quad |V_\mrm{ cb}| &= (41.1 \pm 1.3) \cdot 10^{-3} \, ,\notag
\end{alignat*}
which approximately satisfy the unitarity condition $\sum_i V_\mrm{ ij}V_\mrm{ ik}^* = \delta_\mrm{ jk}$ and
$\sum_j V_\mrm{ ij}V_\mrm{ kj}^* = \delta_\mrm{ ik}$. From the values (\ref{eq:PDG_CKM}), we have 
$\sum_\mrm{ u,c,d,s,b} |V_\mrm{ ij}|^2 = 2.024 \pm 0.032$ (\ie\ with a 1.6\% uncertainty, dominated by the
$\Vcs$ value), although in various cases below we will assume exact
CKM unitarity, \ie we will take $\sum_\mrm{ u,c,d,s,b} |V_\mrm{ ij}|^2 \equiv 2$.
Table~\ref{tab:1} lists the partial and total hadronic widths obtained with and without assuming CKM unitarity. 
The results are compared  (bottom rows) to the values of Ref.~\cite{KARA} obtained for zero quark masses, 
using the 2013 PDG SM input parameters, and without full determination of the associated uncertainties. 
Our result, without imposing CKM unitarity, is lower by about 30~MeV compared to that in~\cite{KARA}, 
mostly due to the updated PDG parameters (the most important are the changes in $\Vcs$ and $|V_{\rm cd}|$ which result in width variations
of $-28$ and $-1.6$~MeV respectively), whereas the inclusion of finite quark masses results in less than a $\sim$1~MeV decrease of the width.
\begin{table}[h!]
\begin{center}
\begin{tabular}{l|c|cccccc|c}\hline\hline
\toprule
Partial widths (MeV)  &  \hspace{0.25cm} $\Gamma^{(0)}$ \hspace{0.5cm} &  \hspace{0.25cm}
$\Gamma^{(1)}_\mrm{QCD}$  \hspace{0.25cm} &  \hspace{0.25cm} $\Gamma^{(2)}_\mrm{QCD}$  \hspace{0.25cm} & \hspace{0.25cm}
$\Gamma^{(3)}_\mrm{QCD}$  \hspace{0.25cm} &  \hspace{0.25cm} $\Gamma^{(4)}_\mrm{QCD}$  \hspace{0.25cm} &
\hspace{0.25cm} $\Gamma_\mrm{ewk}$  \hspace{0.25cm} &  \hspace{0.25cm} $\Gamma_\mrm{mixed}$ & $\GWh$
\hspace{0.25cm} \\ \hline
\midrule
$W \rightarrow q q'$ (exp. $V_\mrm{ ij}$) & 1379.851 & 52.931 & 2.857 & $-$0.992 & $-$0.238 & $-$5.002 & $-$0.755 & 
1428.65 $\pm$ 22.40$_\mrm{ par}$ $\pm 0.04_\mrm{ th}$\\
$W \rightarrow q q'$ ($V_\mrm{ ij}V_\mrm{ jk} = \delta_\mrm{ ik}$) & 1363.197 & 52.291 & 2.822 & $-$0.980 & $-$0.235 & $-$4.942 & $-$0.746 &
1411.40 $\pm~0.96_\mrm{par} \pm 0.04_\mrm{th}$\\\hline 
$W \rightarrow q q'$ (exp. $V_\mrm{ ij}$)~\cite{KARA} & 
1408.980 & 54.087 & 2.927 & $-$1.018 &$-$0.245 & $-$5.132 & $-$0.779 & 1458.820 $\pm~0.006_\mrm{th}$ \\ 
$W \rightarrow q q'$ ($V_\mrm{ ij}V_\mrm{ jk} = \delta_\mrm{ ik}$)~\cite{KARA} & 
1363.640 & 52.346 & 2.833 & $-$0.985 & $-$0.237 & $-$4.940 & $-$0.748 & 1411.910 $\pm~0.006_\mrm{th}$ \\
\hline\hline
\bottomrule
\end{tabular}
\end{center}
\caption{Numerical values (in MeV) of the partial and total hadronic W decay widths computed at N$^3$LO in
  this work, using the experimental CKM matrix  or imposing CKM unitarity, including associated parametric and theoretical 
  uncertainties. The bottom rows show, for comparison, the previous results of~\cite{KARA} (with only partial theoretical uncertainties  
  from missing higher-order terms).} 
\label{tab:1}
\end{table}

\noindent

\begin{figure}[htpb!]
\centering
\includegraphics[width=0.49\textwidth]{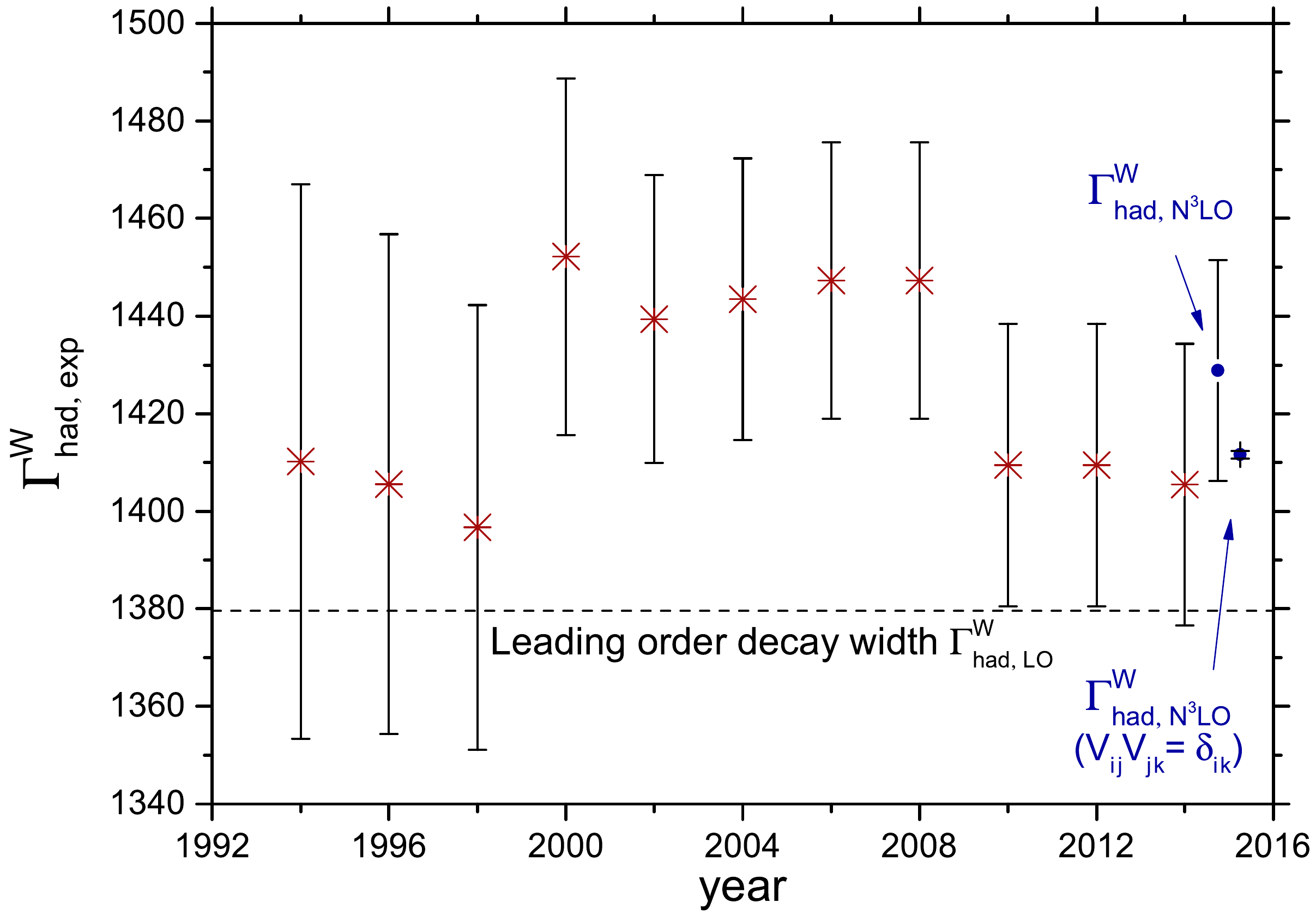}
\includegraphics[width=0.49\textwidth]{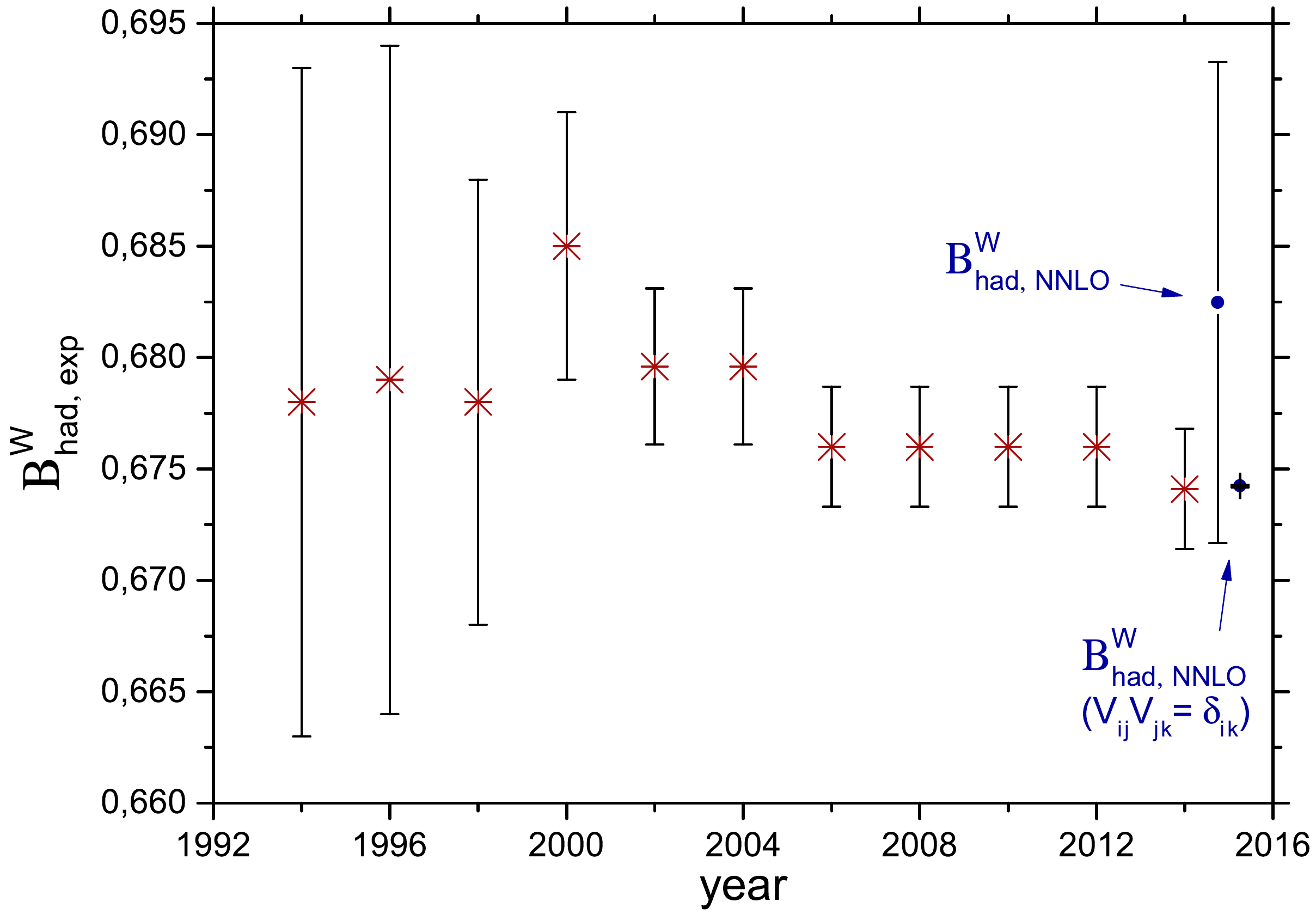}
\caption{Time evolution of the experimental PDG world-average values (red stars)~\cite{PDG} of the hadronic W
  decay width (left) and branching ratio (right) compared to the theoretical predictions 
  computed here with and without imposing CKM unitarity.} 
\label{fig:Gamma_plot_year}
\end{figure}

Our computed W hadronic width, listed in the last column of~Table~\ref{tab:1}, includes two type of uncertainties. The
first ``parametric'' one, clearly dominant, is associated with the uncertainties of the various input parameters
used in the calculations (mostly $\Vcs$, $\MW$, and $\alphas$). The second  ``theoretical'' one is due to
uncertainties mostly from missing higher-order corrections.
The parametric uncertainties have been determined as follows. For each parameter $p=|V_\mrm{ ij}|,\MW,\alphas,...$ 
we have calculated the decay width for $p$, $p+\Delta p$ and $p-\Delta p$, while all other parameters are kept
fixed at their central values. The error on the width is then determined by 
\begin{eqnarray}
\Delta^p_+ \GWh \ &=&  \max \{\GWh(p+\Delta p),\GWh(p),\GWh(p-\Delta p)\} - \GWh(p),\nonumber\\
\Delta^p_- \GWh &=& \GWh (p) -\min \{\GWh(p+\Delta p),\GWh(p),\GWh(p-\Delta p)\}.
\end{eqnarray}
The total parametric errors have been obtained by adding in quadrature the parametric errors from the N parameter variations.
The dominant parametric uncertainty is due to the $\Vcs$ quark coupling strength, whose relative uncertainty of
$1.6\%$~\cite{PDG} propagates into $\pm$22~MeV in $\GWh$. If one assumes CKM unitarity (or, equivalently,
negligible $|V_\mrm{ ij}|$ uncertainties) the second most important source of parametric uncertainty is that from
$\MW$ which propagates into $\pm$0.7~MeV in $\GWh$. 
The theoretical uncertainties of our calculations are clearly much smaller than the parametric ones. 
They are obtained from the quadratic sum of missing higher-order QCD corrections, considered to be of the same
size, $\pm0.019\MeV$, as the $\mathcal{O}(\alphas^5)$ corrections assessed for the hadronic Z boson width~\cite{BCK}; 
plus missing higher-order electroweak and electroweak+QCD terms estimated to be $\pm0.012\MeV$ and
$\pm0.029\MeV$ based on~\cite{KARA}. Non-perturbative effects --suppressed by $\cO{\lqcd^4/\MW^4}$ power
corrections--, zero quark mass approximations beyond LO~\cite{Chetyrkin:1996hm} --estimated to be
$\mathcal{O}(m_\mrm{q}^2/\MW^2)$ and amounting to $\pm0.001 \MeV$ at $\mathcal{O}(\alphas^2)$ and
$\pm0.002\MeV$ at $\mathcal{O}(\alpha)$--, as well as residual effects due to the dependence on the CKM matrix
renormalization scheme --evaluated in~\cite{Almasy:2008ep}--, are much smaller and neglected here. 
In Fig.~\ref{fig:Gamma_plot_year} (left), we compare the yearly evolution of the experimental PDG
world-average $\GWhexp$ (red stars) to the Born $\GWh\approx$~1380~MeV value (dashed line) and to the N$^3$LO
theoretical widths listed in Table~\ref{tab:1}.
Our theoretical results, $\GWh$~=~$1428.65 \pm 22.40_\mrm{par} \pm 0.04_\mrm{th}$~MeV (using the
experimentally measured $|V_\mrm{ij}|$ values), and $\GWh$~=~$1411.40 \pm 0.96_\mrm{par} \pm
0.04_\mrm{th}$~MeV (assuming CKM matrix unitarity), are well in agreement with the current experimental 
value of $\GWhexp = \GWtotexp \cdot \BRWhexp$~=~1405~$\pm$~29~MeV~\cite{PDG} (Table~\ref{tab:summary}).

\vspace{-0.5cm}
\section{Hadronic W branching ratio at NNLO accuracy}

The W hadronic branching fraction, given by the ratio of hadronic to all W decays, is a very simple and
robust experimental observable. It is as inclusive as the total W cross section measureable in \pp\ or $\epem$
collisions but much free from experimental (\eg\ normalization) uncertainties. We obtain its theoretical
numerical value from the ratio $\BRWh = \GWh/\GWtot$, where $\GWh$ is the value computed in the previous
Section, and the total decay width is that obtained from the NNLO calculation of~\cite{ZFITTER} as
parametrized 
in~\cite{Cho:2011rk}.
Using the input parameters (\ref{eq:num_input})--(\ref{eq:PDG_CKM}) and the same procedure to compute parametric and theoretical
uncertainties as for $\GWh$, we obtain $\GWtot = 2093.4 \pm 1.2_\mrm{par} \pm 0.8_\mrm{th} \MeV$, which
agrees well with the experimental value, $\GWtotexp$~=~2085~$\pm$~42~MeV~\cite{PDG}, as well as with the
indirect determination from the full electroweak fit $\GWtotewk$~=~2091~$\pm$~1~MeV~\cite{Baak:2014ora}.
The theoretical NNLO hadronic branching ratio amounts thus to $\BRWh$~=~$0.682 \pm 0.011_\mrm{ par}$ (using the experimental
CKM matrix), with negligible theoretical compared to parametric uncertainties, 
and $\BRWh$~=~$0.6742 \pm 0.0002_\mrm{th} \pm 0.0001_\mrm{par}$ (assuming CKM matrix unitarity).
Note also that the $\MW$ parametric uncertainty cancels out in the $\BRWh$ ratio of hadronic to total W widths.
Both results are in very good accord with the experimental 
value of $\BRWhexp$~=~0.6741~$\pm$~0.0027, as shown in the right plot of Fig.~\ref{fig:Gamma_plot_year} and in Table~\ref{tab:summary}. 



\begin{table}[htbp!]
\centering
\begin{tabular}{l | c  c | c}\hline\hline
	\toprule
         Observable & (full calculation) &  ($V_\mrm{ ij}V_\mrm{ jk} = \delta_\mrm{ ik}$)  & \hspace{0.25cm} Experimental value \hspace{0.25cm}\\ \hline
	\midrule
	$\GWh$ (MeV)   \hspace{0.25cm} &\hspace{0.25cm} $1428.65 \pm 22.40_\mrm{ par} \pm 0.04_\mrm{ th}$ \hspace{0.25cm}&\hspace{0.25cm}  1411.40 $\pm~0.96_\mrm{par} \pm 0.04_\mrm{th}$\hspace{0.25cm} & 1405~$\pm$~29~\\
	$\GWtot$ (MeV) \hspace{0.25cm} &\hspace{0.25cm} $2093.4 \pm 1.2_\mrm{par} \pm 0.8_\mrm{th}$  \hspace{0.25cm}& -- & 2085~$\pm$~42 \\ 
	$\BRWh$        \hspace{0.25cm} &\hspace{0.25cm} $0.682 \pm 0.011_\mrm{ par}~(\pm 0.0002_\mrm{th})$ \hspace{0.25cm}&\hspace{0.25cm} $0.6742 \pm 0.0002_\mrm{th} \pm 0.0001_\mrm{par}$ \hspace{0.25cm}& 0.6741~$\pm$~0.0027 \\ 
	$\RW$          \hspace{0.25cm} &\hspace{0.25cm} $2.15 \pm 0.11_\mrm{par}~(\pm~0.002_\mrm{th})$ \hspace{0.25cm}&\hspace{0.25cm} $2.069~\pm~0.002_\mrm{th}~\pm~0.001_\mrm{par}$ \hspace{0.25cm}& $2.068 \pm 0.025$ \\ \hline \hline
	\bottomrule
\end{tabular}
\caption{W decay parameters computed in this work: Hadronic decay width $\GWh$, total width $\GWtot$, 
hadronic branching ratio $\BRWh$, and hadronic-to-leptonic ratio $\RW$ with their associated
theoretical and parametric uncertainties (using the full calculation with experimentally-measured $V_\mrm{ ij}$ 
elements where needed, or assuming CKM unitarity); compared to the current experimental world averages (last column).} 
\label{tab:summary}
\end{table}

\section{Extraction of $\alphas$}

\begin{figure}[htbp!]
\minipage{0.5\textwidth}
\includegraphics[width=\linewidth]{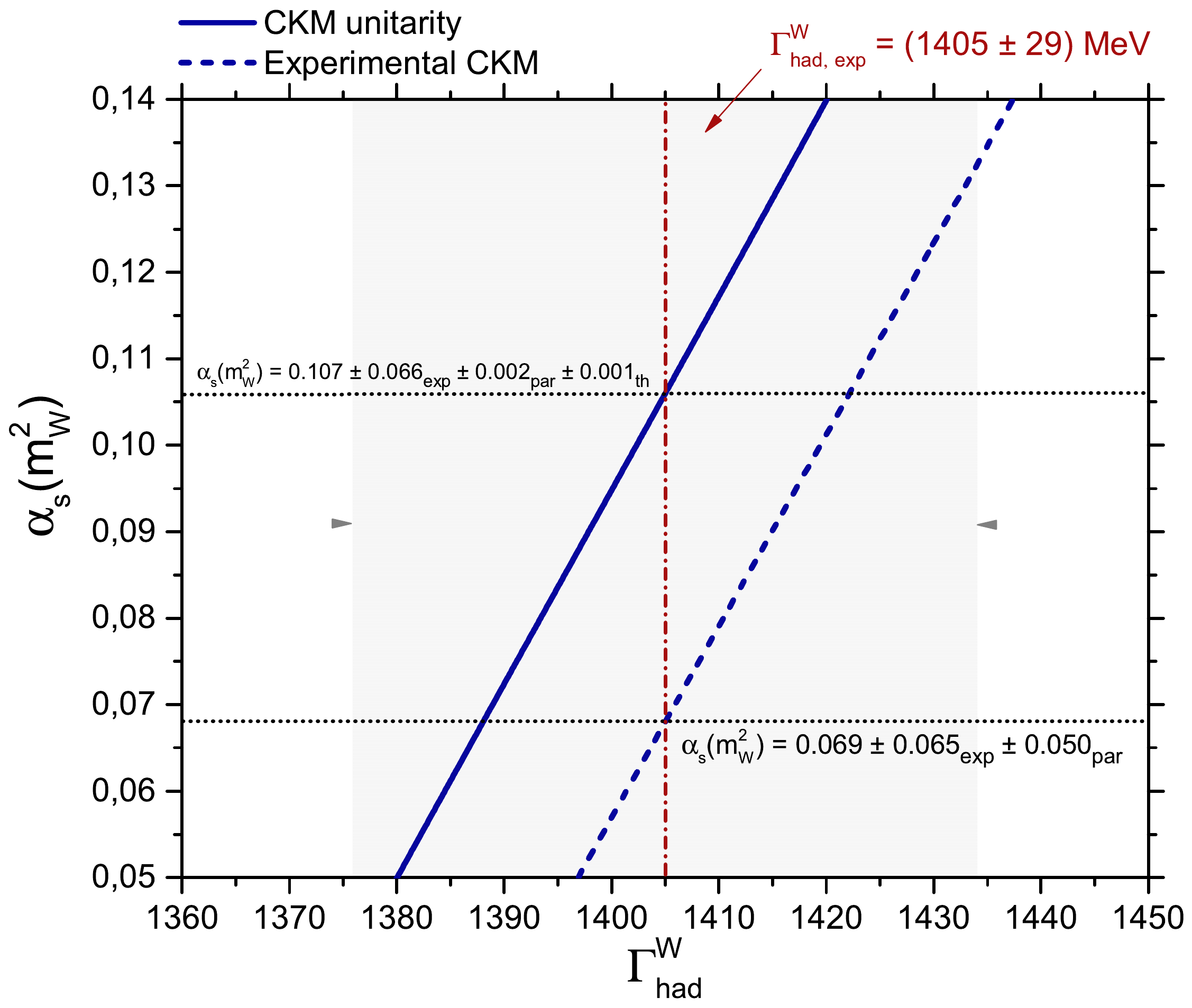}
\endminipage\hfill
\minipage{0.5\textwidth}
\includegraphics[width=\linewidth]{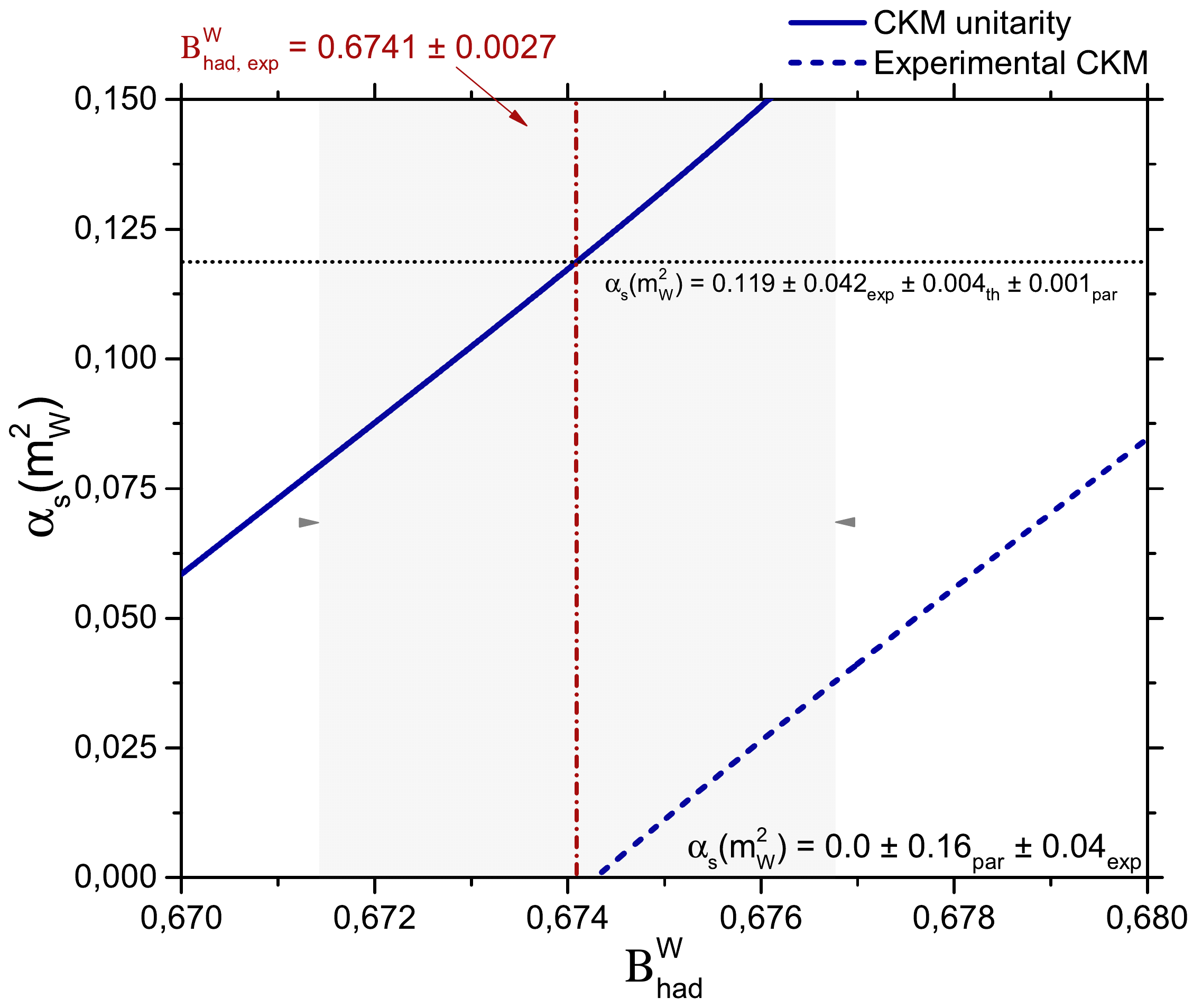}
\endminipage
\caption{Functional dependencies of $\alphas$ on the W hadronic width $\GWh$ (left) and branching ratio
  $\BRWh$ (right), obtained imposing CKM unitarity (solid curves) or using the measured CKM elements (dashed
  curves). The vertical lines are the experimental $\GWhexp$ and $\BRWhexp$ values, and the grey bands
  indicate their current experimental uncertainty.} 
\label{fig:alphas_GWh_BRWh}
\end{figure}

The theoretical dependencies on $\alphas$ of the hadronic W decay width and branching fraction are shown in
Fig.~\ref{fig:alphas_GWh_BRWh} imposing CKM unitarity (solid curves) or using the measured values of the
CKM elements (dashed curves). The vertical lines indicate the current experimental values for both quantities
while the grey bands indicate their associated uncertainties.
Fixing all SM parameters except $\alphas$ to their PDG values, and equating the theoretical expressions for
$\GWh(\alphas)$ and $\BRWh(\alphas)$ to their corresponding experimental measurements, the strong coupling 
can be extracted. The corresponding results are listed in the top rows of Table~\ref{tab:alphas_GWh_BRWh}, where
the obtained $\alphasmW$ values (second column) are evolved to the Z scale (last column) with the NNLO running
coupling expression. As expected, the much larger uncertainty of $\GWh$ ($\pm$2\%) compared to $\BRWh$ ($\pm$0.4\%)
results in a more precise $\alphas$ extraction from the latter. Yet, the current
experimental and parametric uncertainties on $\GWh$ and $\BRWh$ propagate into very large $\alphas$
uncertainties in both cases. Clearly, those results call first for higher precision measurements of 
$\GWtot$ and $\BRWh$. Indicatively, for each MeV of reduced uncertainty on $\GWhexp$ the precision of the
extracted $\alphas$ value would improve by approximately 2\%. 
Secondly, a competitive extraction of $\alphas$ requires also a reduction of the parametric uncertainties of
the calculations. The impact of measuring $\Vcs$ with better precision can be seen by comparing the $\alphas$
values extracted with and without assuming CKM unitarity. Having $\Vcs$ measured with a 
precision comparable to that of $|V_\mrm{ ud}|$ today, namely 5$\cdot$10$^{-4}$, 
would 
make of $\MW$ the leading source of parametric uncertainty on the $\alphas$ value extracted from W hadronic decays.

\begin{table}[htbp!]
\begin{center}
\begin{tabular}{l | c | c}\hline\hline
	$\alphas$ extraction method & $\alphasmW$ & $\alphasmZ$ \\
	\hline
	$\GWh$ (experimental CKM) \hspace{0.25cm} & $0.069 \pm 0.065_\mrm{exp} \pm 0.050_\mrm{par}$ &  $0.068 \pm 0.064_\mrm{exp} \pm 0.050_\mrm{par}$ \\
	$\GWh$ (CKM unitarity) & $0.107 \pm 0.066_\mrm{exp} \pm 0.002_\mrm{par} \pm 0.001_\mrm{th}$ & $0.105 \pm 0.065_\mrm{exp} \pm 0.002_\mrm{par} \pm 0.001_\mrm{th}$ \\\hline
	$\BRWh$ (experimental CKM) & $0.0 \pm 0.04_\mrm{exp} \pm 0.16_\mrm{par}$ & $0.0 \pm 0.04_\mrm{exp} \pm 0.16_\mrm{par}$ \\
	$\BRWh$ (CKM unitarity) & $0.119 \pm 0.042_\mrm{exp} \pm 0.004_\mrm{th} \pm 0.001_\mrm{par}$ & $0.117 \pm 0.042_\mrm{exp} \pm 0.004_\mrm{th} \pm 0.001_\mrm{par}$ \\\hline
 	$\RW$ (experimental CKM) & \hspace{0.25cm} $0.0 \pm 0.04_\mrm{exp} \pm 0.16_\mrm{par}$ & $0.0 \pm 0.04_\mrm{exp} \pm 0.16_\mrm{par}$ \\
 	$\RW$ (CKM unitarity) & \hspace{0.25cm} $0.119 \pm 0.042_\mrm{exp} \pm 0.004_\mrm{th} \pm 0.001_\mrm{par}$ \hspace{0.25cm} & \hspace{0.25cm} $0.117 \pm 0.042_\mrm{exp} \pm 0.004_\mrm{th} \pm 0.001_\mrm{par}$ \hspace{0.25cm}\\\hline\hline
\end{tabular}
\caption{Values of $\alphas$ (and propagated experimental and parametric uncertainties) at the W and Z scales,
extracted from $\GWh$ (top), $\BRWh$ (middle), and $\RW$ (bottom); 
by setting the CKM matrix to the experimental values or imposing CKM unitarity.}
\label{tab:alphas_GWh_BRWh}
\end{center}
\end{table}

\begin{figure}[htbp!]
\includegraphics[width=0.6\linewidth]{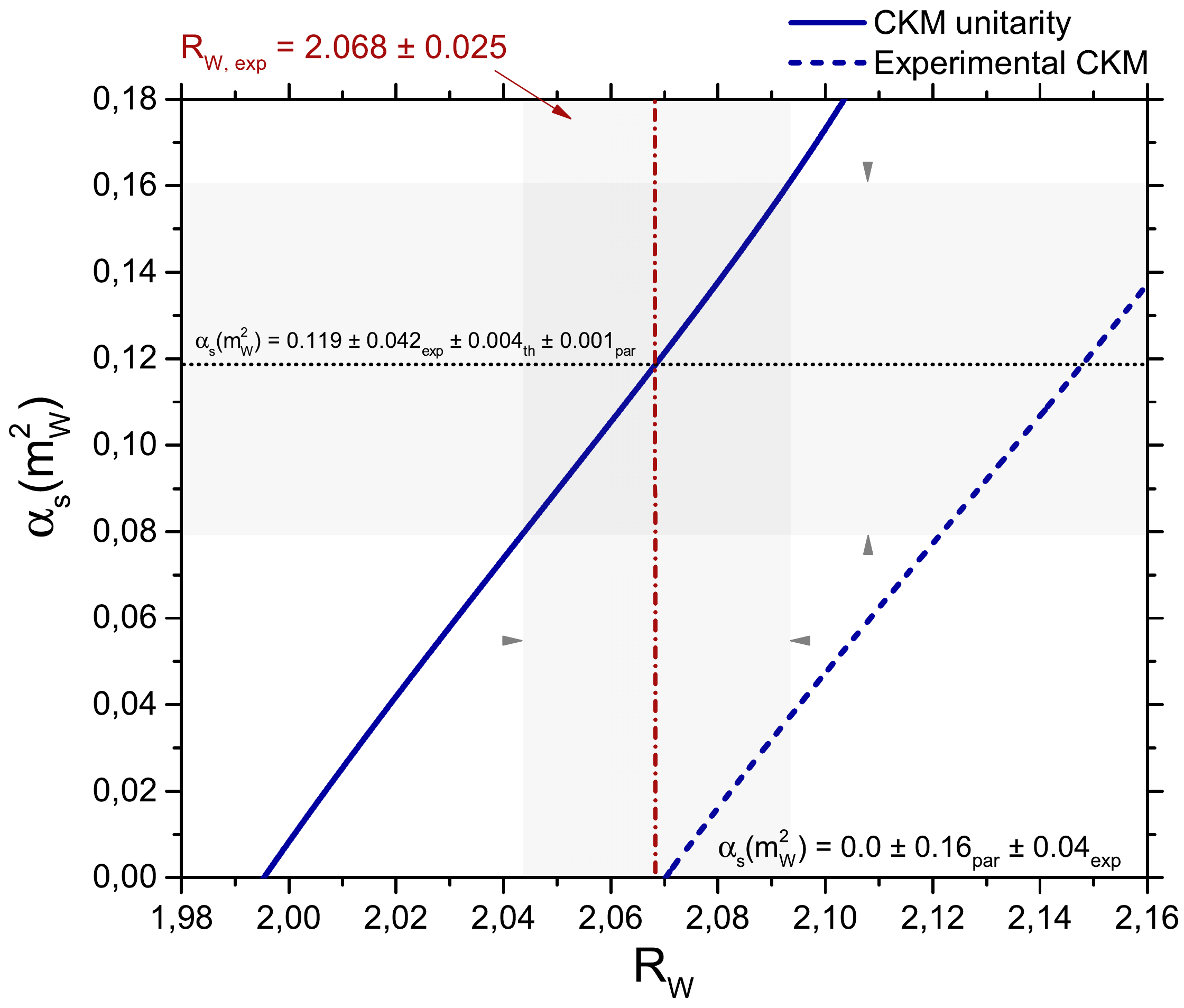}
\caption{Functional dependence of $\alphas$ on the ratio $\RW = \BRWh/\BRWl$, obtained imposing 
  CKM unitarity (solid curves) or using
  the measured CKM elements (dashed curves). The vertical line is the experimental $\RWexp$
  value, and the grey bands indicate its current experimental uncertainty.} 
\label{fig:alphas_RW}
\end{figure}

The experimental values of the leptonic W width ($\GWlexp = 679 \pm 15$~MeV) and branching ratio 
($\BRWlexp = 0.3258 \pm 0.0027$)~\cite{PDG} can also be used to impose constraints on
$\alphas$ through the equalities $\GWh \equiv \GWtot - \GWl$ and $\BRWh \equiv 1 - \BRWl$.
As a matter of fact, the current world values of $\GWh$ and $\BRWh$ have been obtained using also the 
leptonic W decay information~\cite{PDG}. Eventually, for independent high-precision measurements of $\GWhl$ and/or $\BRWhl$
the most efficient way to exploit all experimental information available 
is through the ratio $\RW\equiv\BRWh/\BRWl = \BRWh/(1-\BRWh)$, as done for the Z boson at LEP~\cite{alphas2015}. 
The theoretical $\RW(\alphas)$ dependence
is shown in Fig.~\ref{fig:alphas_RW}, as obtained imposing CKM unitarity (solid curve) or using experimental
CKM elements (dashed curve). The theoretical predictions for the hadronic-to-leptonic W branching ratio 
are $\RW~=~2.069~\pm~0.002_\mrm{th}~\pm~0.001_\mrm{par}$ (assuming CKM unitarity) and $\RW = 2.15 \pm 0.11_\mrm{par}$ 
(experimental CKM), in very good agreement with the empirical result: $\RWexp = 2.068 \pm 0.025$.
The corresponding derived values of $\alphas$ are listed in the bottom rows of Table~\ref{tab:alphas_GWh_BRWh}. 
The final most precise extraction of the QCD coupling from W decays is $\alphasmZ = 0.117 \pm 0.042_\mrm{exp} \pm 0.004_\mrm{th} \pm
0.001_\mrm{par}$, with a relative uncertainty of 35\%, obtained from $\RW$ imposing CKM unitarity.

\section{Extraction of $\Vcs$, and CKM matrix unitarity test}

The hadronic W width, Eq.~(\ref{eq:Gamma_Whad_mq}), involves a sum over the first two rows of the CKM matrix,
\ie the six CKM elements involving quarks lighter than $\MW$ listed in (\ref{eq:PDG_CKM}).
Among these, the $|V_\mrm{ ud}|$ and $\Vcs$ terms are the most important in W hadronic decays and,
as shown previously, the least precisely known ($\Vcs$) contributes to the largest uncertainty in the
calculation of $\GWh$ and $\BRWh$. 
From the theoretical expressions and the experimental values of  the hadronic width and branching ratio,
fixing all SM parameters to their world-averages except $\Vcs$, we can extract the charm-strange mixing parameter. 
The corresponding results are listed in the middle column of Table~\ref{tab:Vcs}. The associated experimental,
parametric, and theoretical $\Vcs$ uncertainties are propagated as explained before for the $\alphas$
determination. The $\GWhexp$ and $\BRWhexp$ uncertainties propagate into $\pm$2\% and $\pm$0.4\% respectively,
the parametric uncertainties are of order $\pm$0.2\%, and the theoretical ones are negligible
($\pm0.0004_\mrm{th}$) and not quoted. Our most precise extraction, combining hadronic and leptonic branching 
fractions through the $\RW$ ratio, yields $\Vcs = 0.973 \pm 0.004_\mrm{exp} \pm 0.002_\mrm{par}$, with a 0.5\% uncertainty,
improving by a factor of four the precision of the current world-average experimental value,  $\Vcsexp = 0.986
\pm 0.016$~\cite{PDG}. As a matter of fact, the W decays provide the most stringent test of CKM unitarity
today. Indeed, leaving free the sum $\sum_\mrm{ u,c,d,s,b} |V_\mrm{ ij}|^2$ in the theoretical expression for
$\BRWh$, the hadronic-to-leptonic ratio measurement of $\RWexp = 2.069 \pm 0.018$ implies
$\sum_\mrm{ u,c,d,s,b} |V_\mrm{ ij}|^2 = 1.999 \pm 0.008_\mrm{exp} \pm 0.001_\mrm{th}$ 
(with negligible $\pm 0.0002_\mrm{par}$ parametric uncertainty).


\begin{table}[htbp!]
\centering
\begin{tabular}{l | c | c }\hline\hline
	\toprule
	Extraction method \hspace{0.25cm} & $\Vcs$ & $\sum_\mrm{ u,c,d,s,b}|V_\mrm{ ij}|^2$ \\ \hline
	\midrule
	$\GWh$  & $0.969 \pm 0.021_\mrm{exp} \pm 0.002_\mrm{par} $ & \hspace{0.25cm} $1.991 \pm 0.041_\mrm{exp} \pm 0.001_\mrm{par} \pm 0.001_\mrm{th}$ \hspace{0.2cm} \\
	$\BRWh$ & $0.973 \pm 0.004_\mrm{exp} \pm 0.002_\mrm{par} $ &  $1.999 \pm 0.008_\mrm{exp} \pm 0.001_\mrm{th}$ \\ 
	$\RW$ & \hspace{0.25cm} $0.973 \pm 0.004_\mrm{exp} \pm 0.002_\mrm{par}$ \hspace{0.25cm} & \hspace{0.25cm} $1.999 \pm 0.008_\mrm{exp} \pm 0.001_\mrm{th}$ \\\hline 
	Experimental value & \hspace{0.25cm} $0.986 \pm 0.016$ \hspace{0.25cm} & $2.024 \pm 0.032$  \\\hline \hline
	\bottomrule
\end{tabular}
\caption{Values of the charm-strange CKM element $\Vcs$ (second column), and sum of the first six CKM matrix
  elements squared $\sum_\mrm{ u,c,d,s,b}|V_\mrm{ ij}|^2$ (last column), with their propagated uncertainties,
  extracted from different experimental W decay observables; compared to their experimental values (bottom row).} 
\label{tab:Vcs}
\end{table}


\vspace{-0.35cm}
\section{Future prospects}

\begin{figure}[!htb]
\minipage{0.5\textwidth}
\includegraphics[width=\linewidth]{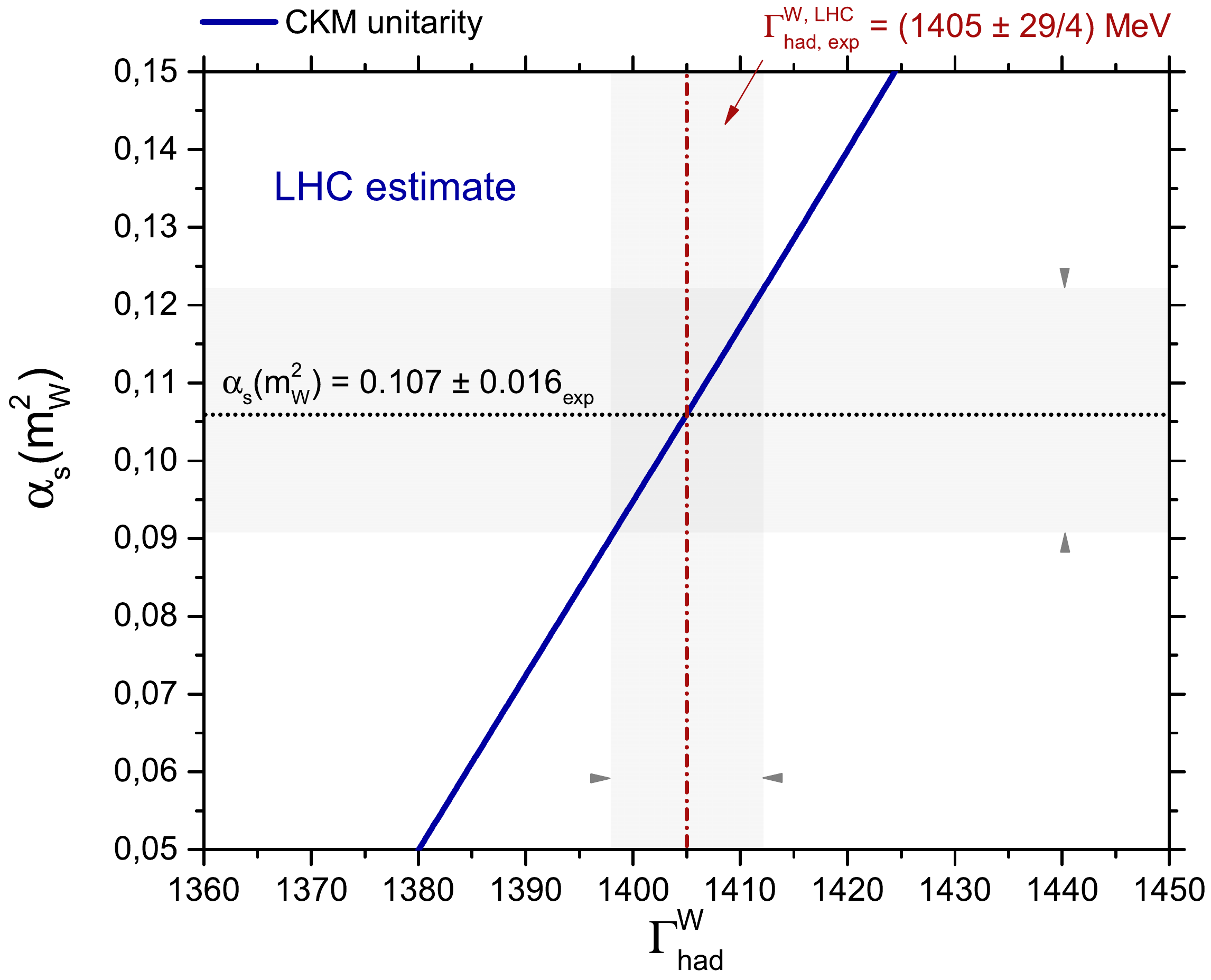}
\endminipage\hfill
\minipage{0.5\textwidth}
\includegraphics[width=\linewidth]{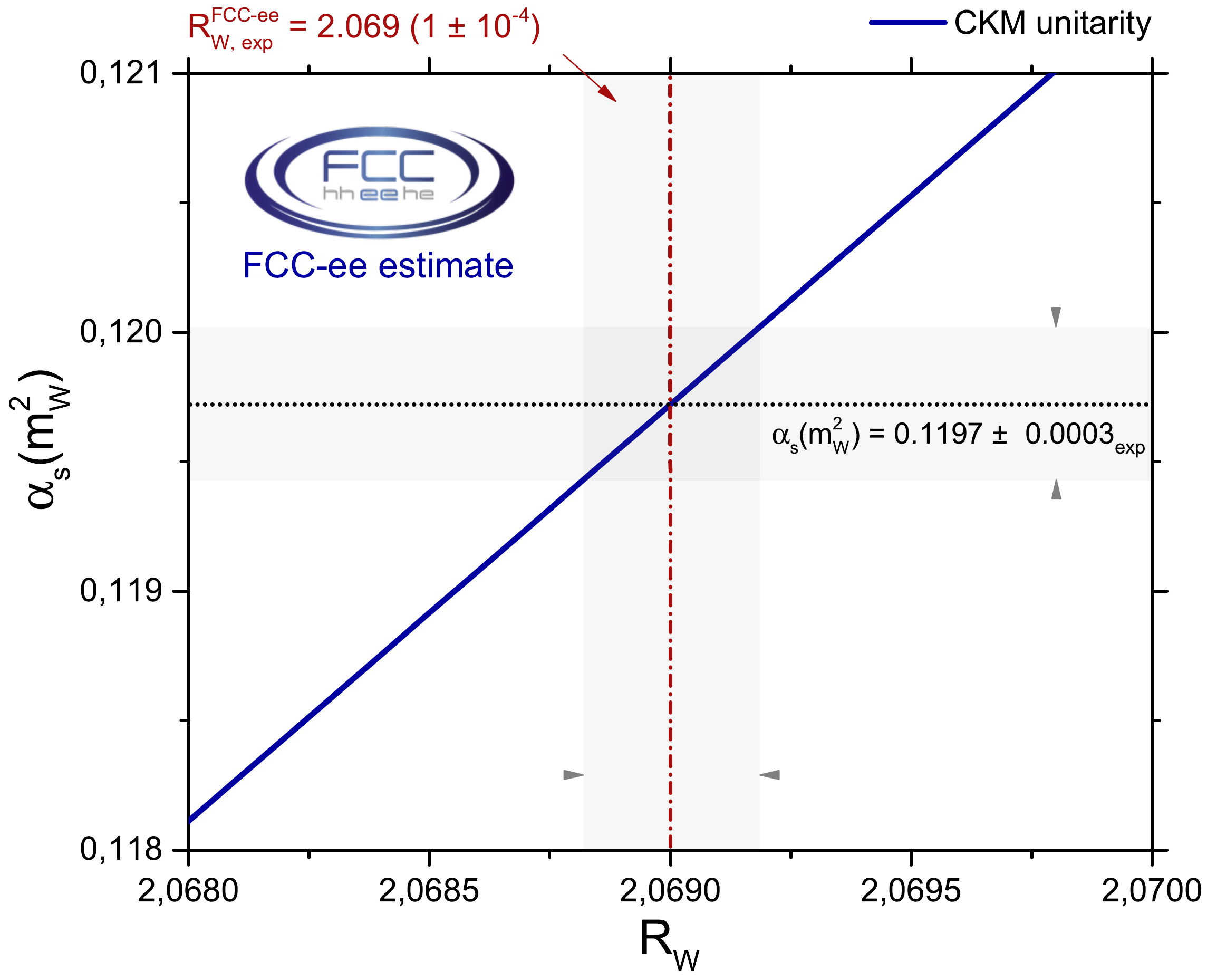}
\endminipage
\caption{Estimated future extractions of $\alphas$ from the W hadronic width $\GWh$ (left) and from the hadronic/leptonic decay ratio $\RW$
  (right). The vertical lines are the current experimental $\GWhexp$ and $\RWexp$ central values with the
  horizontal grey bands indicating the expected future experimental uncertainties on $\alphas$ at the LHC (left),
  and at the FCC-ee (right).}  
\label{fig:alphas_GWh_BRWh_future}
\end{figure}

A precise determination of the strong coupling, as well as stringent SM tests such as CKM unitarity,
require measurements of W decays of higher precision than those available today. 
The total W width has been directly measured 
via maximum-likelihood fits of (i) the Breit-Wigner W mass distribution in  $\epem\to\,$W$^{+}$W$^{-}$, yielding
$\GWtotexp$~=~2195~$\pm$~83~MeV~\cite{Schael:2013ita}, as well as of (ii) the tail of the W transverse mass
$m_{_\mrm{ T}}(\ell\nu)$ spectrum in leptonic W$\to\ell\nu$ decays in \pp,~\ppbar~$\to\rm W+X$ collisions, yielding
$\GWtotexp$~=~2046~$\pm$~49~MeV~\cite{TEW:2010aj} (their combination yielding the experimental world average 
quoted in Table~\ref{tab:summary}). The branching fraction $\GWh$ can only be measured with 
small uncertainties in $\epem\to\,$W$^{+}$W$^{-}$~\cite{Schael:2013ita}, although a competitive $\GWh = 1 - \GWl$ value 
can be obtained from precise measurements of the total W width and the leptonic branching ratio exploiting the large W data 
samples at \pp,~\ppbar\ colliders~\cite{TEW:2010aj,Camarda:2016twt}.
Measurements at the LHC and future $\epem$ colliders will provide $\GWh$, $\BRWh$ and $\RW$ with higher accuracy and precision. 
In the hadron collider determinations of $\GWtot$ and $\BRWl$, the leading source of systematic uncertainties comes from the 
proton parton distributions functions (PDF), amounting to 70\% and 60\% respectively~\cite{TEW:2010aj,Camarda:2016twt}.
At the LHC, a maximum factor of four reduction of the current uncertainties on the derived value of $\BRWhexp$ 
can be assumed thanks to our improved knowledge of PDFs, and the much higher statistics available in measurements of the large-$m_{_\mrm{ T}}(\ell\nu)$ 
spectra (Fig.~\ref{fig:alphas_GWh_BRWh_future}, left). Combining all upcoming W decays measurements at the LHC 
with the currently available results, can thereby reduce the propagated $\alphas$ experimental uncertainty 
to the 10\% level, but going below this can only be achieved through high-precision $\epem$ measurements.
In $\epem\to\rm W^{+}W^{-}$ at the FCC-ee, the total W width $\GWtot$ can be accurately measured through a
threshold scan around $\sqrts = 2\MW$, and also the W hadronic branching ratio $\BRWh$ would profit from the
huge sample of $5 \times 10^8$ W bosons (a thousand times more than the $5 \times 10^5$ W's collected at
LEP)~\cite{tlep} which would reduce the statistical uncertainty of $\BRWh$ to around 0.005\%. Thus, neglecting
parametric uncertainties, a $\BRWh$ measurement at the FCC-ee would significantly improve the extraction of
$\alphas$ with propagated experimental uncertainties of order 0.4\%. The $\alphas$ uncertainty could be
further lowered down to $\sim$0.2\% 
through the measurement of the $\RW$ ratio in three $\epem\to\rm W^{+}W^{-}$ final states,
such as $\rm \ell\nu\,\ell\nu,\;\ell\nu\,qq,\;qq\,qq$. Indeed, the ratio of cross sections $\sigma(\rm WW \to qq\;qq)/\sigma(WW \to
\ell\nu\;\ell\nu)$ is proportional to $(R_\mrm{W})^2$, thereby gaining a factor two in statistical sensitivity,
and being totally independent of potential modifications of the weak coupling running and  
free from cross section normalization uncertainties~\cite{tlep}. Figure~\ref{fig:alphas_GWh_BRWh_future} shows
the estimated $\alphas$ extractions from the expected improved measurements of $\GWh$ alone at the LHC (left), and $\RW$ at FCC-ee (right).


\section{Summary}

To summarize, we have calculated the numerical values of the hadronic W decay width ($\GWh$) and its hadronic
branching ratio ($\BRWh$) at N$^3$LO and NNLO accuracy respectively, improving upon previous theoretical
results and carefully estimating the associated experimental, theoretical and parametric uncertainties. The
computed values  
$\GWh$~=~$1428.65 \pm 22.40_\mrm{par} \pm 0.04_\mrm{th}$~MeV and
$\BRWh$~=~$0.682 \pm 0.011_\mrm{ par}$ (using the experimental
CKM matrix elements), and $\GWh$~=~$1411.40 \pm 0.96_\mrm{par} \pm 0.04_\mrm{th}$~MeV 
and $\BRWh$~=~$0.6742 \pm 0.0002_\mrm{th} \pm 0.0001_\mrm{par}$ (assuming CKM matrix unitarity), 
are in very good agreement with the corresponding experimental measurements: $\GWhexp$~=~1405~$\pm$~29~MeV and
$\BRWhexp$~=~0.6741~$\pm$~0.0027. Also the obtained ratios of hadronic-to-leptonic branching fractions, 
$\RW = 2.069 \pm 0.002_\mrm{th} \pm 0.001_\mrm{par}$ (assuming CKM unitarity) and $\RW = 2.15 \pm 0.11_\mrm{par}$ 
(experimental CKM elements), are in very good agreement with the measured value $\RWexp = 2.068 \pm 0.025$.
By comparing the experimental results to the theoretical expectations, 
we have extracted the strong coupling $\alphas$, and the charm-strange CKM element $\Vcs$ under different
assumptions. The current experimental and parametric uncertainties on $\GWh$, $\BRWh$ and $\RW$ are too large
today to allow for a precise determination of $\alphas$ (the best result obtained is 
$\alphasmZ = 0.117 \pm 0.042_\mrm{exp} \pm 0.004_\mrm{th} \pm 0.001_\mrm{par}$, assuming CKM matrix unitarity) although
upcoming high-statistics W measurements at the LHC could reduce the $\alphas$ extraction uncertainties to the
$\sim$10\% level. Our study shows that a future high-luminosity $\epem$ collider such as FCC-ee running at 
$\sqrts\approx 2\MW$ will allow for an $\alphas$ determination with uncertainties as low as 0.2\%.\\

We have also quantified the constraints that the hadronic W decays impose on the quark mixing parameters as
encoded in the CKM matrix of the Standard Model. By fixing all SM parameters, including $\alphas$, to their
default values and leaving free $\Vcs$ in the theoretical expressions for $\BRWh$, we can determine 
the charm-strange coupling with a 0.5\% uncertainty, $\Vcs = 0.973 \pm 0.004_\mrm{exp} \pm 0.002_\mrm{par}$,
which is four times better than the current world-average experimental value, $\Vcsexp = 0.986 \pm 0.016$. Similarly,
the experimental values of the hadronic and leptonic W branching fractions imply 
$\sum_\mrm{ u,c,d,s,b} |V_\mrm{ ij}|^2 = 1.999 \pm 0.008_\mrm{exp} \pm 0.001_\mrm{th}$, providing today the
most stringent test of CKM unitarity for the five lightest quarks.\\

\paragraph*{\bf Acknowledgments.} We are grateful to A.~Blondel, P.~Janot, J.~F.~Kamenik, D. Kara, B.~Kniehl, D.~Nomura,
and T.~Riemann for useful discussions. M. Srebre is grateful to financial support from the CERN EP/TH
Departments where part of this work was carried out.


\end{document}